\documentclass{ifacconf}

\usepackage{graphicx}      
\usepackage{natbib}        

\usepackage{setspace}
\usepackage{amsmath}
\usepackage{caption}
\usepackage{amssymb}
\newtheorem{assumption}{Assumption}
\newtheorem{theorem}{Theorem}
\def\salt{\vskip 0.2 true cm}
\def\cl{\mathcal}
\def\udef{\buildrel \bigtriangleup \over =}
\def\nbeq#1{(\ref{#1})}

\def\UNION{\mathop{\bigcup}}
\def\qed{\hfill $\square$}

\begin{document}
\begin{frontmatter}

\title{Distributed Adaptive Fault-Tolerant Control of Uncertain Multi-Agent Systems}

\author[First]{Mohsen Khalili} 
\author[Second]{Xiaodong Zhang}
\author[Third]{Marios Polycarpou} 
\author[Fourth]{Thomas Parisini}
\author[Fifth]{Yongcan Cao}  

\address[First]{Wright State University, 
   Dayton, OH 45435, USA (e-mail: khalili.4@wright.edu).}
\address[Second]{Wright State University, 
   Dayton, OH 45435, USA (e-mail: xiaodong.zhang@wright.edu).}
\address[Third]{ Department of Electrical
and Computer Engineering, University of Cyprus, Nicosia, Cyprus (e-mail: mpolycar@ucy.ac.cy).}
\address[Fourth]{Imperial College London and University of Trieste (e-mail: t.parisini@gmail.com).}
\address[Fifth]{Air Force Research Laboratory, Wright-Patterson AFB, OH 45435, USA (e-mail: yongcan.cao@gmail.com).}

\begin{abstract}
This paper presents an adaptive fault-tolerant control (FTC) scheme for a class of nonlinear uncertain multi-agent systems. A local FTC scheme is designed for each agent using local measurements and suitable information exchanged between neighboring agents. Each local FTC scheme consists of a fault diagnosis module and a reconfigurable controller module comprised of a baseline controller and two adaptive fault-tolerant controllers activated after fault detection and after fault isolation, respectively. Under certain assumptions, the closed-loop system's stability and leader-follower consensus properties are rigorously established under different modes of behavior of the FTC system, including the time-period before possible fault detection, between fault detection and possible isolation, and after fault isolation.
\end{abstract}


\end{frontmatter}

\section{Introduction}
 Several modern technical systems can be characterized by distributed multi-agent systems, that is, systems comprised of various distributed and interconnected autonomous agents/subsystems. Examples of such systems include cooperative unmanned vehicles, smart grids, air traffic control system, etc. In recent years, cooperative control using distributed consensus algorithms has received significant attention~(see, e.g., \cite{Ren2008}).  Since the overall distributed multi-agent systems are required to operate reliably at all times, despite the possible occurrence of faulty behaviors in some agents, the development of fault diagnosis and accommodation schemes is a crucial step in achieving reliable and safe operations.

In the last two decades, significant research progress has been made in the design and analysis of fault diagnosis and accommodation schemes~(see, for instance, \cite{Blanke:book2006}). Most of these methods utilize a centralized architecture, where the diagnostic module is designed based on a global mathematical model of the overall system and is required to have real-time access to all sensor measurements. Because of limitations of computational resource and communication overhead, such centralized methods may not be suitable for large-scale distributed interconnected systems. As a result, in recent years, there has been a significantly increasing research interest in distributed fault diagnosis schemes for multi-agent systems (see, for instance,
\cite{YanEdwards:IJC2008,Ferrari:TAC2012,Johansson:Automatica2011}).

This paper presents a distributed adaptive FTC methodology for accommodating faults in a class of nonlinear uncertain multi-agent systems. A FTC scheme is designed for each agent in the distributed system by utilizing local measurements and suitable information exchanged between neighboring agents. Each local FTC scheme consists of two main modules: 1) the online health monitoring (fault diagnosis) module consists of a bank of nonlinear adaptive estimators. One of them is the fault detection estimator, while the others are fault isolation estimators; and 2) the reconfigurable controller (fault accommodation) module consists of a baseline controller and two adaptive fault-tolerant controllers used after fault detection and after fault isolation, respectively. Under certain assumptions, the closed-loop system's stability and leader-following consensus properties are established for the baseline controller and adaptive fault-tolerant controllers. This paper significantly extends the results of (\cite{Zhang:TAC2004}) by generalizing the centralized FTC method to the case of leader-follower formation of distributed multi-agent systems.

\section{Graph Theory Notation}
A directed graph $\cl{G}$ is a pair $(\cl{V},\cl{E})$, where $\cl{V} = \{v_1,\cdots,v_P \}$ is
a set of nodes, $\cl{E} \subseteq \cl{V}\times\cl{V}$ is a set of edges, and $P$ is the number of nodes. An edge is an ordered pair of distinct nodes $(v_j,v_i)$ meaning that the $i$th node can receive information from $j$th node.
For an edge $(v_j,v_i)$, node $v_j$ is called the parent node, node $v_i$ the child node, and $v_j$ is a neighbor of $v_i$. An undirected graph can be considered as a special case of a directed graph where $(v_i,v_j) \in \cl{E}$ implies $(v_j,v_i) \in \cl{E}$ for any $v_i, v_j \in \cl{V}$. An undirected graph is connected if there is a path between any pair of nodes. A directed graph contains a directed spanning tree if there exists a node called the root such that the node has directed paths to all other nodes in the graph. 

The set of neighbors of node $\upsilon_i$ is denoted by $N_i = \{j : (\upsilon_j,\upsilon_i) \in \cl{E}\}$.
The weighted adjacency matrix $\cl{A} = [a_{ij}] \in \Re^{P\times P}$ associated with the directed graph $\cl{G}$ is defined by $a_{ii} = 0$, $a_{ij} > 0$ if $(\upsilon_j, \upsilon_i) \in \cl{E}$, and $a_{ij} = 0$ otherwise. The topology of an intercommunication graph $\cl{G}$ is said to be fixed, if each node has a fixed neighbor set and $a_{ij}$ is fixed. It is clear that for undirected graphs $a_{ij}=a_{ji}$. The Laplacian matrix $L = [ l_{ij} ] \in \Re^{P \times P}$ is defined as $l_{ii} = \sum_{j\in N_i} a_{ij}$ and $l_{ij} = - a_{ij}$, $i\neq j$. Both $\cl{A}$ and $L$ are symmetric for undirected graphs and $L$ is positive semidefinite. 

\section{Problem Formulation} \label{sec:Problem}
Consider a set of $M$ agents with the dynamics of the $i$th agent, $i=1,\cdots,M$, being described by
\begin{eqnarray}
\begin{array} {ccl}
\dot x_i &=& \phi_i(x_i) + u_i(x_i,x_J) + \eta_i(x_i,t) \\
&&+ \beta_i(t-T_i) f_i(x_i,u_i(x_i,x_J)),  
\end{array}
\label{eq:sys-model}
\end{eqnarray}
where $\, x_i \in \Re^{n} \, $ and  $\, u_i \in \Re^{n} \,$ are the state vector and input vector of the $i$th agent, respectively. Additionally, $x_J$ contains the state variables of neighboring agents that directly communicate with agent $i$, including the time-varying leader to be tracked (i.e., $x^r$) as agent number $M+1$, i.e., $J=\{j:j\in N_i\}$,
$\, \phi_i: \Re^{n} \mapsto \Re^{n}$, $\,\eta_i : \Re^{n} \times \Re^{+} \mapsto \Re^{n} \,$ and $f_i: \Re^n \times \Re^n \mapsto \Re^{n}$ are smooth vector fields. Specifically, $\phi_i$ and $\eta_i$
represent the known nonlinearity and modeling uncertainty, respectively.
The term $\beta_i(t-T_i) f_i(x_i,u_i)$ denotes the changes in the dynamics of $i$th agent due to the occurrence of a fault. Specifically, $\beta_i(t-T_i)$ represents the time profile of a fault which occurs at some unknown time $T_i$, and $f_i(x_i,u_i)$ is a nonlinear fault function. In this paper, $\beta_{i}(\cdot)$ is assumed to be a step function (i.e., $\beta_{i}( t-T_i) =0$ if $ t<T_i$, and  $\beta_{i} (t-T_i) =1$ if $ t\ge T_{i}$). It is assumed that in each agent only one fault possibly occurs at any time.

\noindent {\bf Remark 1}: The distributed multi-agent system model given by (\ref{eq:sys-model}) is a nonlinear generalization of the single integrator dynamics considered in literature (for instance, \cite{Ren2008}). In this paper, in order to investigate the fault-tolerance and robustness properties, the fault function $f_i(x_i)$ and modeling uncertainty $\eta_i$ are included in the system model.


For isolation purposes, we assume that there are $r_i$ types of possible nonlinear fault functions in the fault class associated with the $i$th agent; specifically, $f_i(x_i,u_i)$ belongs to a finite set of functions given by
\begin{equation}
\cl{F}_i \udef \{f_i^1 (x_i,u_i),\cdots,f_i^{r_i} (x_i,u_i)\}\,.
\label{eq:problem:F}
\end{equation}
Each fault function $f_i^s$, $s=1,\cdots,r_i$, is described by
\begin{equation}
f_i^s (x_i,u_i) \udef [\big(\theta_{i1}^s\big)^T \hspace{-1mm}g_{i1}^s(x_i,u_i),\cdots,\big(\theta_{in}^s\big)^T \hspace{-1mm}g_{in}^s(x_i,u_i)]^T ,
\label{eq:faultclass}
\end{equation}
 where $\theta_{ip}^s$, for $i=1,\cdots,M$, and $p=1,\cdots,n$, is an unknown parameter assumed to belong to a known compact set $\Theta_{ip}^s$ (i.e., $\theta_{ip}^s \in \Theta_{ip}^s \subseteq \Re^{q_{ip}^s}$), and $g_{ip}^s: \Re^n \times \Re^n\mapsto \Re^{q_{ip}^s}$ is a known smooth vector field. As described in \cite{Zhang:TAC2004}, the fault model described by (\ref{eq:problem:F}) and (\ref{eq:faultclass}) characterizes a general class of nonlinear faults where the vector field $g_{ip}^s$ represents the functional structure of the $s$th fault affecting the $p$th state equation, while the unknown parameter vector $\theta_{ip}^s$ characterizes the fault magnitude.

The objective of this paper is to develop a robust distributed fault-tolerant leader-following consensus control scheme, using diagnostic information, for the class of distributed multi-agent systems described by  \nbeq{eq:sys-model}. The following assumptions are made throughout the paper:

\salt
\begin{assumption} \label{Assump:uncertainty}
Each component of the modeling uncertainty, represented by $\eta_{i}(x_{i},t)$ in (\ref{eq:sys-model}), has a known upper bound, i.e., $\forall p=1,\cdots,n$,
 $ \forall x_i \,\in\, \Re^n$ and $ \forall u_i \,\in \Re^n $,
\begin{equation}
| \eta_{ip}(x_{i},t) | \leq \,\bar{\eta}_{ip} (x_{i},t) \,,
\label{eq:uncertainty:bound1} 
\end{equation}
where the bounding function $\, \bar{\eta}_{ip} \,$ is known and uniformly bounded.
\end{assumption}
\begin{assumption} \label{Assump:intercommunicationGraph}
The intercommunication topology of the distributed system described by (\ref{eq:sys-model}) is a fixed connected undirected graph.
\end{assumption}
Assumption~\ref{Assump:uncertainty} characterizes the class of modeling uncertainty under consideration. The bound on the modeling uncertainty is needed in order to distinguish between the effects of faults and modeling uncertainty during the fault diagnosis process.
Assumption~\ref{Assump:intercommunicationGraph} is needed to ensure that the information exchange among agents is sufficient for the team to achieve the desired team goal.

%
Let us define three important
time--instants: $T_i$ is the fault occurrence time; $\,
T_d > T_i \,$ is the time--instant when a fault is detected; $\, T_{\rm isol}
> T_d \,$ is the time--instant when the monitoring system
(possibly) provides a fault isolation decision, that is, which
fault in the class $\, {\cl F}_i \,$ has actually occurred. The structure of
the fault-tolerant controller for the $i$th agent takes on the following general form (\cite{Zhang:TAC2004}): 
\begin{equation}
\begin{array}{rcl}
\dot \omega_i &=& \left \{
\begin{array}{ll}
g_0 (\omega_i ,x_i,x_J,t) \, ,  \quad     &  {\rm for} \quad t < T_d \\
g_D (\omega_i ,x_i,x_J,t) \, , \quad     & {\rm for} \quad T_d \le t < T_{\rm isol} \\
g_I (\omega_i ,x_i,x_J,t) \, , \quad     & {\rm for} \quad t \ge T_{\rm isol} \\
\end{array} \right . \\
\\
u_i &=& \left \{
\begin{array}{ll}
h_0 (\omega_i ,x_i,x_J,t) \, , \quad & {\rm for} \quad t < T_d \\
h_D (\omega_i ,x_i,x_J,t) \, , \quad & {\rm for} \quad T_d \le t < T_{\rm isol} \\
h_I (\omega_i ,x_i,x_J,t) \, , \quad & {\rm for} \quad t \ge T_{\rm isol} \\
\end{array}  \right .
\end{array}
\label{eq:general:contr}
\end{equation}
where $ \omega_i $ is the state vector of the distributed controller;
  $\, g_0, g_D,
g_I $ and $\, h_0, h_D, h_I $ are nonlinear functions to be designed
according to the following qualitative objectives:
\begin{enumerate}
\item
In a fault free mode of operation, a baseline controller guarantees the state of $i$th agent $x_i(t)$ should track the leader's time-varying state $x^r$, even in the possible presence of plant modeling uncertainty. 
%

\item If a fault is detected by diagnostic scheme, the baseline controller is reconfigured to compensate for the effect of the (yet unknown) fault, that is, the fault-tolerant controller is designed in such a way as to exploit the information that a fault has occurred, so that the controller may recover some control performance. This new controller should guarantee the boundedness of system signals and some leader-following consensus performance, even in the presence of the fault.  

\item If the fault is isolated by diagnostic scheme, then the controller is reconfigured again. The second fault-tolerant controller is designed using the information about the type of fault that has actually occurred so as to improve the control performance.

\end{enumerate}

\section{Baseline Controller Design}
In this section, we design the baseline controller and investigate the closed-loop system stability and performance before fault occurrence.
Without loss of generality, let the leader be agent number $M+1$
with a time-varying reference state (i.e., $x_{M+1}=x^r$). The baseline controller for the $i$th agent can be designed as:
\begin{equation}
u_{ip} = -\sum_{j\in N_i} k_{ij}\tilde x_{ij} -\phi_{ip}(x_{ip}) - \bar \kappa_{ip}\, sgn\big(\sum_{j \in N_i} k_{ij} \tilde x_{ij}\big) ,
\label{eq:nominal:controller}
\end{equation}
where $u_{ip}$ and $x_{ip}$ are the $p$th component of the input and state vectors of the $i$th agent, respectively, $p=1,\cdots,n$, $i = 1,\cdots,M$, $\tilde{x}_{ij} \udef x_{ip}-x_{jp}$, $\bar \kappa_{ip} \udef \bar \eta_{ip}+\kappa_p$, $\kappa_p$ is a positive bound on $|\dot{x}_p^r|$ (i.e., $\kappa_p \geq |\dot{x}_p^r|$), $sgn(\cdot)$ is the sign function, $N_i$ is the set of neighboring agents that directly communicate with the $i$th agent including the leader, and $k_{ij}$, for $j\in N_i$, are positive constants. Notice that $k_{im}=0$, for $m\notin N_i$.

Note that, by considering the leader as agent $M+1$, the topology graph for the $M+1$ agents has a spanning tree with the leader as its root. First, we need the following Lemmas:
\begin{lem} (\cite{Ren2008}) \label{lem:laplacian}
The Laplacian matrix $L\in \Re^{P\times P}$ of a directed graph $\cl G$ has at least one 0 eigenvalue with $\textbf{1}_P$ as its right eigenvector, where $\textbf{1}_P$ is a $P\times 1$ column vector of ones, and all nonzero eigenvalues of $L$ have positive real parts. 0 is a simple eigenvalue of $L$ if and only if the directed graph $\cl G$ has a spanning tree.
\newline
\end{lem}
\begin{lem} \label{lem:laplaciansquared}
Consider a connected graph $\cl{G}$ with the leader as the $(M+1)$th node.
The matrix
\begin{equation}
\bar{\cl{L}} \udef \Psi \cl{L}+\cl{L}^T \Psi
\label{Lbar}
\end{equation}
is positive semidefinite and has a simple zero eigenvalue with $\textbf{1}_{M+1}$ as its right eigenvector, where $\Psi\in \Re^{(M+1)\times (M+1)}$ is the Laplacian matrix of the graph with an undirected leader,
and $\cl{L}\in \Re^{(M+1)\times (M+1)}$ is the Laplacian matrix of the graph with a directed leader.
\end{lem}
\begin{pf}
Decomposing the Laplacian matrices $\cl{L}$ and $\Psi$, we have
\begin{eqnarray} \label{eq:Decomposition}
\cl{L} = \begin{bmatrix}
L_{11} & L_{12} \\
\\
0_{1\times M} & 0_{1\times 1}
\end{bmatrix}, \hspace{3mm}
\Psi = \begin{bmatrix}
L_{11} & L_{12} \\
\\
L_{12}^T & L_{22}
\end{bmatrix}\,,
\end{eqnarray}
where $L_{11}\in \Re^{M\times M}$ is a symmetric matrix, $L_{12}\in \Re^{M\times 1}$, and $L_{22}\in \Re$. Based on Lemma~\ref{lem:laplacian}, the matrix $\cl{L}$ is a positive semidefinite matrix having a simple zero eigenvalue with $\mathbf{1}_{M+1}$ as its right eigenvector. Therefore,
the specific structure of $\cl{L}$ defined in (\ref{eq:Decomposition}) implies that $L_{11}$ is a positive definite matrix, having all its eigenvalues in the right-hand plane. From (\ref{eq:Decomposition}), we obtain
\begin{equation*}
\bar{\cl{L}} = \Psi \cl{L}+\cl{L}^T \Psi = \begin{bmatrix}
2L_{11}^2 & 2L_{11}L_{12} \\
2L_{12}^T L_{11} & 2L_{12}^T L_{12}
\end{bmatrix}\,.
\end{equation*}
Let $\chi$ be the eigenvalue of $\bar{\cl{L}}$. We have
\begin{equation*}
|\chi I_{M+1}-\bar{\cl{L}}| = \begin{vmatrix}
\chi I_M-2L_{11}^2 & -2L_{11}L_{12} \\
-2L_{12}^T L_{11} & \chi-2L_{12}^T L_{12}
\end{vmatrix}\,,
\end{equation*}
where $I$ represents the identity matrix.

Using $\begin{vmatrix}
A & B \\
C & D
\end{vmatrix}=|A|\cdot|D-C\,A^{-1}B|$, we have
\begin{eqnarray*}
\lefteqn{ |\chi I_{M+1}-\bar{\cl{L}}| = |\chi I_M-2L_{11}^2|} \\
&&\cdot|\chi-2L_{12}^T L_{12} - 4L_{12}^T L_{11}(\chi I_M-2L_{11}^2)^{-1} L_{11}L_{12}|\,.
\end{eqnarray*}
The eigenvalues of $\bar{\cl{L}}$ satisfying $|\chi I_M-2L_{11}^2|=0$ are all positive, because $L_{11}$ and therefore $2L_{11}^2$ are positive definite. Furthermore, $\chi=0$ satisfies $|\chi-2L_{12}^T L_{12}-4L_{12}^T L_{11}(\chi I_M-2L_{11}^2)^{-1} L_{11}L_{12}|=0$. Additionally, it can be shown $\bar{\cl{L}} \mathbf{1}_{M+1}=\Psi \cl{L} \mathbf{1}_{M+1}+\cl{L}^T \Psi \mathbf{1}_{M+1} = 0$. Therefore, the proof of Lemma~\ref{lem:laplaciansquared} can be concluded.
\qed
\end{pf}
\noindent {\bf Remark 2}: It is worth noting that the Laplacian matrix $\Psi$ for the undirected graph is only considered for the purpose of controller performance analysis. The actual distributed control topology is directed, since the leader is only sending the data and does not receive any data from other agents.

The following result characterizes the stability and leader-following performance properties of the controlled system before fault occurrence.
\begin{theorem} \label{thm:nominal}
In the absence of faults in the $i$th agent, the baseline controller described by \nbeq{eq:nominal:controller} 
guarantees that the leader-follower consensus is achieved asymptotically with a time-varying reference state, i.e. $x_i-x^r\rightarrow 0 $ as $t \rightarrow \infty$.
\end{theorem}
\begin{pf}
Based on (\ref{eq:nominal:controller}) and before occurrence of the fault, the closed-loop system dynamics are given by
 \begin{equation} \label{eq:baseline:outputdynamics}
\dot x_{ip} = -\sum_{j\in N_i}k_{ij} \tilde{x}_{ij}+ \eta_{ip} - (\bar \eta_{ip}+\kappa_p) sgn\big(\sum_{j \in N_i} k_{ij} \tilde{x}_{ij} \big)\,,
\end{equation}
We can represent the collective state dynamics as
\begin{equation}
\dot{x}^p = -\cl{L} x^p + \zeta^p - \bar{\zeta}^p \,,
\label{eq:nominal:outputdynamics}
\end{equation}
where $x^p \in \Re^{M+1}$ is comprised of the $p$th state component of the $M+1$ agents, including the leader as the $(M+1)$th agent, i.e., $x^p=\begin{bmatrix}
x_{1p}, & x_{2p}, & \cdots, & x_{Mp}, & x^r_p\,
\end{bmatrix}^T$, the terms $\zeta^p \in \Re^{M+1}$ and $\bar{\zeta}^p \in \Re^{M+1}$ are defined as
\begin{equation}
\label{eq:eta_p}
\hspace{-0.5mm}\begin{array}{ccl}\zeta^p \udef \begin{bmatrix} \eta_{1p}, & \cdots, & \eta_{Mp}, & 0
\end{bmatrix}^T \end{array} \hspace{-1mm},
\begin{array}{ccl}\bar{\zeta}^p \udef  \begin{bmatrix}
\bar \zeta_{1p}, & \cdots, & \bar \zeta_{Mp}, &
0
\end{bmatrix}^T \end{array}\hspace{-1mm}, 
\end{equation}
where $\bar \zeta_{ip} \udef (\bar \eta_{ip}+\kappa_p) sgn\big(\sum_{j \in N_i} k_{ij} \tilde{x}_{ij} \big)$, $i=1,\cdots,M$.
We consider the following Lyapunov function candidate:
\begin{equation}
\begin{array}{ccl}
V_p = {x^p}^T \Psi x^p = \frac{1}{2} \sum_{i=1}^{M} \sum_{j\in N_i} k_{ij} \tilde x_{ij}^2 + \frac{1}{2} \sum_{i=1}^M k_{i(M+1)} \tilde x_{i(M+1)}^2 ,
\end{array}
\label{eq:nominal:V}
\end{equation}
where $\Psi$ is defined in Lemma~\ref{lem:laplaciansquared}, $\tilde x_{i(M+1)} \udef x_{ip}-x_{(M+1)p}$, and $x_{(M+1)p}$ is the $p$th component of the leader's time-varying state $x^r$.
Then, the time derivative of the Lyapunov function (\ref{eq:nominal:V}) along the solution of (\ref{eq:nominal:outputdynamics}) is given by
\begin{eqnarray}
\dot{V}_p &=& -{x^p}^T \bar{\cl{L}} \, {x^p} + 2\dot{x}_p^r \sum_{i=1}^M k_{i(M+1)} (x_p^r-x_{ip}) \nonumber\\
&& + 2{x^p}^T \Psi (\zeta^p-\bar{\zeta}^p)\,,
\label{eq:Nominal:Vdot1}
\end{eqnarray}
where $\bar{\cl{L}}$ is defined in (\ref{Lbar}). Based on (\ref{eq:eta_p}), we have
\begin{eqnarray}
\label{eq:Nominal:Vdot2}
\hspace{-5mm}{x^p}^T \Psi \zeta^p &=& \sum_{i=1}^M \sum_{j\in N_i} k_{ij} (x_{ip}-x_{jp}) \eta_{ip} \,, \\ \label{eq:Nominal:Vdot3}
\hspace{-5mm}{x^p}^T \Psi \bar{\zeta}^p &=& \sum_{i=1}^M \sum_{j\in N_i} k_{ij} \tilde{x}_{ij} (\bar \eta_{ip}+\kappa_p) sgn\big(\sum_{j \in N_i} k_{ij} \tilde{x}_{ij}\big).
\end{eqnarray}
Using the property that $k_{ij} = k_{ji}$ (based on Assumption~\ref{Assump:intercommunicationGraph}), we know that $\sum_{i=1}^M \sum_{j\in N_i, j\neq M+1} k_{ij} (x_{ip}-x_{jp})=0$. Therefore, we have
\begin{equation}
2\dot{x}_p^r \sum_{i=1}^M k_{i(M+1)} (x_p^r-x_{ip}) = 
-2 \dot{x}_p^r \sum_{i=1}^M \sum_{j\in N_i} k_{ij} \tilde{x}_{ij}.
\label{eq:Nominal:Vdot3_4}
\end{equation}
By substituting (\ref{eq:Nominal:Vdot2}), (\ref{eq:Nominal:Vdot3}) and (\ref{eq:Nominal:Vdot3_4}) into (\ref{eq:Nominal:Vdot1}), we have
\begin{eqnarray}
\hspace{-5mm}\dot{V}_p &=& -{x^p}^T \bar{\cl{L}} \, {x^p} + 2\sum_{i=1}^M \sum_{j\in N_i} k_{ij} \tilde{x}_{ij} (\eta_{ip}-\dot{x}_p^r) \nonumber\\
&& - 2\sum_{i=1}^M \sum_{j\in N_i} k_{ij} \tilde{x}_{ij} (\bar \eta_{ip}+\kappa_{p}) sgn\bigg(\sum_{j \in N_i} k_{ij} \tilde{x}_{ij}\bigg).
\label{eq:baseline:Vdot4}
\end{eqnarray}
Based on (\ref{eq:baseline:Vdot4}) and Assumption~\ref{Assump:uncertainty}, we have
\begin{equation*}
\dot{V}_{p} \leq -{x^p}^T \bar{\cl{L}} \, {x^p}\,.
\end{equation*}
Therefore, using Lemma~\ref{lem:laplaciansquared}, we know that $\dot{V}_p $ is negative definite with respect to $x_{ip}-x_{jp}$, because the only $x^p$ that makes $-{x^p}^T \bar{\cl{L}} \, {x^p}$ zero is $x^p=\mathbf{1}_{M+1} c$, where $c$ is a constant. Therefore, consensus is reached asymptotically, i.e., $x_{ip}-x_{jp}\rightarrow 0$ as $t\rightarrow \infty$. More specifically, $x_{ip}-x^r_p\rightarrow 0$ as $t\rightarrow \infty$.
\qed
\end{pf}

\section{Distributed Fault Diagnosis} \label{sec:Diagnosis}
The distributed fault detection and isolation (FDI) architecture is comprised of $M$
local  FDI  components,  with  one  FDI  component
designed for each of the $M$ agents. The objective
of each local FDI component is to detect and isolate faults
in the corresponding agent. Specifically, each
local  FDI  component  consists  of  a  fault  detection
estimator (FDE) and a bank of $r_i$ nonlinear adaptive
fault isolation estimators (FIEs), where $r_i$ is the number of different nonlinear
fault  types  in  the  fault  set $\mathcal{F}_i$ (\ref{eq:problem:F}) associated  with  the corresponding agent. Under normal  conditions,  each  local  FDE  monitors  the
corresponding local agent to detect the occurrence
of  any  fault.  
If  a  fault  is  detected  in  a  particular
agent $i$, then the corresponding $r_i$ local FIEs are
activated for the purpose of determining the particular
type of fault that has occurred in the agent.
The FDI design for each agent follows
the  generalized  observer  scheme  architectural  framework (\cite{Blanke:book2006}).
The distributed FDI algorithm is designed by extending the centralized algorithm in \cite{Zhang:TAC2004}. 
  
\renewcommand{\thesubsection}{\arabic{section}.\arabic{subsection}}
\subsection{Distributed Fault Detection} \label{subsec:Detection}
Based on the agent model described by (\ref{eq:sys-model}), the FDE for each agent is chosen as:
\begin{eqnarray}
\begin{array}{ccl}
\dot {\hat x}_i &=& \phi_i(x_i)+u_i + H_i (x_i-\hat x_i)\,, 
\end{array}
\label{eq:detection:estimator}
\end{eqnarray}
where $\hat{x}_i \in \Re^n$ denote the estimated local state, $H_i= {\rm diag}\{h_{i1},\cdots,h_{in}\}$ is a positive definite matrix, where $-h_{ip}<0$ is the estimator pole, $p=1,\cdots,n$, $i=1,\cdots,M$. Without loss of generality, let the observer gain be $H_i= h_i I_n$ where $I_n$ is a $n\times n$ identity matrix. It is worth noting that
the  distributed  FDE  (\ref{eq:detection:estimator})  for  the  $i$th  agent  is constructed based on local input and state variables (i.e. $u_i$ and $x_i$) and certain communicated information $x_j$ from the FDE associated with the $j$th agent.

For each local FDE, let $\tilde{x}_i \udef x_i-\hat{x}_i$ denote the state estimation error of the $i$th agent. Then, before fault occurrence (i.e., for $0\leq t<T_i$), by using (\ref{eq:sys-model}) and (\ref{eq:detection:estimator}), the estimation error dynamics are given by
\begin{eqnarray}
\begin{array}{ccl}
\dot{\tilde{x}}_i &=& -H_i \tilde{x}_i + \eta_i(x_i,t)\,.
\end{array}
\label{eq:detection:outputerror}
\end{eqnarray}
The presence of modeling uncertainty $\eta_i(x_i,t)$ causes a nonzero estimation error. A bounding function on the state estimation error $\tilde{x}_{ip}$, before the occurrence of the fault can be derived. Specifically, based on Assumption \ref{Assump:uncertainty}, for $0 \leq t < T_i$, each component of the state estimation error $\tilde{x}_{ip}$ satisfies
\begin{eqnarray}
|\tilde{x}_{ip}| &\leq& \int_0^t e^{-h_i(t-\tau)} \, \bar{\eta}_{ip}\, d\tau +\bar{x}_{ip} e^{-h_i t} \nonumber\,,
\label{eq:detection:xthreshold}
\end{eqnarray}
where $\bar{x}_{ip}$ is a conservative bound on the initial state estimation error (i.e., $|\tilde{x}_{ip}(0)|\leq \bar{x}_{ip}$). Therefore, for each component of the state estimation error (i.e., $\epsilon_{ip}=x_{ip}-\hat{x}_{ip}$), by using (\ref{eq:detection:outputerror}) and applying the triangle equality, we have
$|\epsilon_{ip}| \leq \nu_{ip}$,
where
\begin{eqnarray}
\nu_{ip}(t) &\udef& \int_0^t e^{-h_i(t-\tau)} \, \bar{\eta}_{ip}(x_{i},\tau) \, d\tau + \bar{x}_{ip} e^{-h_i t} \,.
\label{eq:detection:bound}
\end{eqnarray}
Note that the integral term in the above threshold can be easily implemented as the output of a linear filter with the input given by $\bar{\eta}_{ip} (x_{i},t)$. 

Thus, we have the following: 

\noindent \textit{Fault Detection Decision Scheme:}  The decision on the occurrence of a fault (detection) in the $i$th agent is made when the modulus of at least one component of the state estimation error (i.e., $\epsilon_{ip}(t)$) generated by the local FDE exceeds its corresponding threshold $\nu_{ip}(t)$ given by (\ref{eq:detection:bound}).

The fault detection time $T_d$ is defined as the first time instant such that $\vert \epsilon_{ip} \vert > \nu_{ip}$, for some $T_d \geq T_i$ and some $p \in \{ 1,\cdots,n \}$, that is, 
\begin{eqnarray*}
T_d \udef {\rm inf} \, \UNION_{p=1}^{n}\{t \geq 0 \,:\, |\epsilon_{ip}(t)|>\nu_{ip}(t) \}
\label{eq:detection:time}
\end{eqnarray*}

\subsection{Distributed Fault Isolation \label{subsec:Isolation}}
Now, assume
that a fault is detected in the $i$th agent at some
time $T_d$; accordingly, at $t = T_d$ the FIEs in the local
FDI  component  designed  for  the  $i$th  agent  are
activated.  Each  local FIE  is  designed  based  on  the  functional structure of one potential fault type in the agent (see (\ref{eq:faultclass})).  Specifically,  the  following  $r_i$ nonlinear
adaptive estimators are designed as isolation estimators:
for $s = 1, \cdots , r_i$ ,
\begin{eqnarray}
\dot {\hat x}_i^s &=& \phi_i(x_i) + u_i + \Lambda_i^s (x_i-\hat x_i^s) + f_{i}^s(x_i,u_i,\hat{\theta}_i^s) \nonumber\\
 f_i^s (x_i,u_i,\hat{\theta}_i^s) &=& [\big(\hat{\theta}_{i1}^s\big)^T g_{i1}^s(x_i,u_i),\cdots,\big(\hat{\theta}_{in}^s\big)^T g_{in}^s(x_i,u_i)]^T\,, \nonumber\\ 
 \label{eq:isolationestimator}
\end{eqnarray}
where $\hat {\theta}_i^s$, for $i=1,\cdots,M$, and $s=1,\cdots,r_i$, is the estimate of the fault parameter vector in the $i$th agent, $\Lambda_i^s$ is a diagonal positive definite matrix. For notational simplicity and without loss of generality, in this paper we assume that $\Lambda_i^s = \lambda_i\, I_n$, for all $s=1,\cdots,r_i$.

The adaptation in the isolation estimators is due to the unknown fault parameter vector $\theta_i^s$. The adaptive law for updating each $\hat{\theta}_i^s$ is derived by using the Lyapunov synthesis approach (\cite{Ioannou:adaptive96}), with the projection operator restricting $\hat{\theta}_i^s$ to the corresponding known set $\Theta_i^s$. Specifically, if we let $\epsilon_{ip}^s(t) = x_{ip}-\hat{x}_{ip}^s$, $p=1,\cdots,n$, be  the $p$th component of the state estimation error generated by the $s$th FIE associated with the $i$th agent, then the following adaptive algorithm is chosen:
\begin{eqnarray*}
\dot{\hat{\theta}}_{ip}^s = \mathcal{P}_{\Theta_{ip}^s}\{\gamma_{ip}^s g_{ip}^s(x_{i},u_i) \epsilon_{ip}^s\}\,,
\label{eq:isolation:adaptivelaw}
\end{eqnarray*}
where $\gamma_{ip}^s>0$ is a constant learning rate.

Based on (\ref{eq:sys-model}) and  (\ref{eq:isolationestimator}), each component of the state  estimation error dynamics in the presence of fault $s$ is given by
\begin{eqnarray*}
\dot{\tilde{x}}_{ip}^s &=& -\lambda_i \tilde{x}_{ip}^s + \eta_{ip}(x_i,t) - \big(\tilde{\theta}_{ip}^s \big)^T g_{ip}^s(x_i,u_i) \,,
\label{eq:isolation:error}
\end{eqnarray*}
where, for $p=1,\cdots,n$, $\tilde{x}_{ip}^s$ is the state estimation error, $ \tilde{\theta}^s_{ip} = \hat{\theta}^s_{ip}-\theta^s_{ip}$ is the parameter estimation error. Therefore, by using the triangle equality, a bound on each component of the state estimation error can be obtained as
\begin{eqnarray*}
|\tilde{x}_{ip}^s|  &\leq & \int_{T_d}^t e^{-\lambda_i(t-\tau)} \bigg[\bar{\eta}_{ip}(x_{i},\tau) + \xi_{ip}^s \, \big|g_{ip}^s(x_{i},u_i)\big|\bigg] d\tau \nonumber\\
&& + \bar{x}_{ip}^s e^{-\lambda_i(t-T_d)} \,,
\label{eq:isolation:xthreshold}
\end{eqnarray*}
where $\bar{x}_{ip}^s$ is a conservative bound on the initial state estimation error (i.e., $|\tilde{x}_{ip}^s(T_d)|\leq \bar{x}_{ip}^s$), and $\xi_{ip}^s$ represents the maximum fault parameter vector estimation error, i.e., $|\theta_{ip}^s-\hat{\theta}_{ip}^s(t)| \leq \xi_{ip}^s$. The form of $\xi_{ip}^s$ depends on the geometric properties of the compact set $\Theta_{ip}^s$ (\cite{Zhang:TAC2004}). For instance, assume that the parameter set $\Theta_{ip}^s$ is a hypersphere (or the smallest hypersphere containing the set of all possible $\hat{\theta}_{ip}^s(t)$ with center $O_{ip}^s$ and radius $R_{ip}^s$); then we have $\xi_{ip}^s = R_{ip}^s + |\hat{\theta}_{ip}^s(t)-O_{ip}^s|$.

Therefore, each component of the state estimation error $\epsilon_{ip}^s$, $p=1,\cdots,n$, satisfies
$|\epsilon_{ip}^s| \leq \mu_{ip}^s$,
where
\begin{eqnarray}
\mu_{ip}^s(t) &=& \int_{T_d}^t e^{-\lambda_i(t-\tau)} \bigg[ \bar{\eta}_{ip}(x_{i},\tau) + \xi_{ip}^s \, \big|g_{ip}^s(x_{i},u_i) \big| \bigg] d\tau \nonumber\\
&& + \bar{x}_{ip}^s \, e^{-\lambda_i(t-T_d)} \,.
\label{eq:isolation:thresholds}
\end{eqnarray}
The fault isolation decision scheme is based on the following intuitive principle: if fault $s$ occurs at some time $T_i$ and is detected at time $T_d$, then a set of threshold functions $\mu_{ip}^s(t)$ can be designed such that each component of the state estimation error generated by the $s$th estimator satisfies $|\epsilon_{ip}^s(t)| \leq \mu_{ip}^s(t)$ for all $t\geq T_d$. In the fault isolation procedure, if for a particular fault isolation estimator $b$, there exists some $p \in\{1,\cdots,n\}$, such that the $p$th component of its state estimation error satisfies $|\epsilon_{ip}^b(t)| > \mu_{ip}^b(t)$ for some finite time $t>T_d$, then the possibility of the occurrence of corresponding fault type can be excluded.
Based on this intuitive  idea,  the  following  fault  isolation  decision scheme is devised.

\noindent \textit{Distributed fault isolation decision scheme:} If for each $b \in \{1, \cdots, r_i \}\backslash \{s\}$, there exist some finite time $t^b>T_d$ and some $p \in \{1, \cdots, n \}$, such that $|\epsilon^b(t^b)| > \mu_{ip}^b(t^b)$, then
the  occurrence  of  fault  $s$  in  the  $i$th  subsystem  is concluded.



\section{Fault-Tolerant Controller Module}
In this section, the design and analysis of the FTC schemes are rigorously investigated for two different operating modes of the closed-loop system: 1) during the period after fault detection and before isolation, and 2) after fault isolation.
To facilitate the analysis of the distributed adaptive FTC systems, from now on we assume that the general fault function $f_i^s (x_i,u_i)$ given in (\ref{eq:faultclass}) takes on the following specific forms:

\begin{enumerate}
\item Process faults represented by
\begin{equation}
 f_i^s (x_i) \udef [\big(\theta_{i1}^s\big)^T g_{i1}^s(x_i),\cdots,\big(\theta_{in}^s\big)^T g_{in}^s(x_i)]^T \,,
 \label{eq:processfaultclass}
\end{equation}
\item Actuator fault represented by partial loss of effectiveness of the actuators. Specifically,
\begin{equation}
 f_i^s (u_i) \udef [\theta_{i1}^s u_{i1},\cdots,\theta_{in}^s u_{in}]^T \,,
 \label{eq:actuatorfaultclass}
\end{equation}
where the parameter $\theta_{ip}^s \in (-1,0]$, $p={1,\cdots,n}$, characterizes the magnitude of the actuator fault.
\end{enumerate}

\subsection{Accommodation before Fault Isolation}
After the fault is detected at time $t=T_d$, the isolation estimators are activated to determine the particular type of fault that has occurred.
Meanwhile, the nominal controller is reconfigured to ensure the system stability and some tracking performance after fault detection. In the following, we describe the design of the fault-tolerant controller using adaptive tracking techniques.
Before the fault is isolated, no information about the fault type and fault function is available. Adaptive approximators such as neural-network models can be used to estimate the unknown process fault function $\beta_i f_i$. The term ``adaptive approximator" (\cite{Polycarpou:book2006}) is used to represent nonlinear multivariable approximation models with adjustable parameters, such as neural networks, fuzzy logic networks, polynomials, spline functions, etc. Specifically, we consider linearly parametrized network (e.g., radial-basis-function networks with fixed centers and variances) described as follows: for $p=1,\cdots,n$,
\begin{equation}
\hat{f}_{ip}(x_i,\hat{\vartheta}_{ip}) = \sum_{j=1}^\varrho c_{pj} \varphi_j(x_i) \,,
\label{eq:fi_hat-Before-Isolation}
\end{equation}
where $\varphi_j(\cdot)$ represents the fixed basis functions, and $\hat{\vartheta}_{ip} \udef col(c_{pj} : j=1,\cdots,\varrho)$ is the adjustable weights of the nonlinear approximator. In the presence of a process fault, $\hat f_{ip}$ provides the adaptive structure for online approximating the unknown fault function $f_{ip}(x_i)$. This is achieved by adapting the weight vector $\hat{\vartheta}_{ip}(t)$. 
Therefore, the system dynamics described by (\ref{eq:sys-model}) can be rewritten as, for $p=1,\cdots,n$,
\begin{equation}
\begin{array}{ccl}
\dot x_{ip} \hspace{-0.2mm}=\hspace{-0.2mm} \phi_{ip}(x_{ip}) + (1+\theta_{ip}) u_{ip} + \eta_{ip} +\hat{f}_{ip}(x_{i},{\vartheta}_{ip}) + \delta_{ip}(x_i) ,
\end{array}
\label{eq:FTC:model_beforeIsolation}
\end{equation}
where the parameter $\theta_{ip}$ is defined in (\ref{eq:actuatorfaultclass}), $\delta_{ip} \udef f_{ip}(x_i)-\hat{f}_{ip}(x_{i},{\vartheta}_{ip})$ is the network approximation error for the $p$th state of the $i$th agent, and $\vartheta_{ip}$ is the optimal weight vector given by
\begin{equation*}
\vartheta_{ip} \udef {\rm{arg}} \underset{\hat{\vartheta}_{ip} \in \Theta_{ip}}{\rm{inf}} \bigg\{ \underset{x_{i}\in\cl{X}_i}{\rm{sup}} |f_{ip}(x_i)-\hat{f}_{ip}(x_{i},\hat{\vartheta}_{ip})|\bigg\}\,,
\end{equation*}
where $\cl{X}_i \subseteq \Re^n$ denotes the set to which the variables $x_i$ belongs for all possible modes of behavior of the controlled system. To simplify the subsequent analysis, in the following we assume that the bounding conditions on the network approximation error are global, so we set $\cl{X}_i=\Re^n$.
For each network, we make the following assumption on the network approximation error:
\begin{assumption} \label{Assump:BoundingFunction}
for each $i=1,\cdots,M$, and $p=1,\cdots,n$,
\begin{equation}
|\delta_{ip}| \leq \alpha_{ip} \bar{\delta}_{ip}(x_i)\,,
\end{equation}
\end{assumption}
where $\bar{\delta}_{ip}$ is a known positive bounding function, and $\alpha_{ip}$ is an unknown constant.

Based on the system model (\ref{eq:FTC:model_beforeIsolation}), the neural network model (\ref{eq:fi_hat-Before-Isolation}), and Assumption~\ref{Assump:BoundingFunction}, an adaptive neural controller can be designed using adaptive approximation and bounding control techniques (\cite{Polycarpou:book2006}). Specifically, we consider the following controller algorithm:
\begin{eqnarray}
\label{eq:FTC:controller_beforeIsolation}
u_{ip} &=& \frac{1}{1+\hat{\theta}_{ip}} \bar{u}_{ip} \,, \\
\label{eq:FTC:controller-bar_beforeIsolation}
\bar u_{ip} &=& -\phi_{ip}(x_{ip})-\sum_{j\in N_i} \big(k_{ij}\tilde{x}_{ij} \big)- \hat{f}_{ip}(x_{i},\hat{\vartheta}_{ip}(t)) -\psi_{ip} \nonumber\\
&& - (\bar \eta_{ip}+\kappa_p) sgn\bigg(\sum_{j \in N_i} k_{ij} \tilde{x}_{ij} \bigg)\,, \\
\label{eq:FTC:adaptive_beforeIsolation}
\dot{\hat{\vartheta}}_{ip} &=& \Gamma_{ip} \sum_{j\in N_i} k_{ij} \tilde{x}_{ij} \varphi_{ip}(x_{i})\,,
\end{eqnarray}
\begin{eqnarray}
\label{eq:FTC:adaptiveterm_beforeIsolation}
\psi_{ip} &=& \hat{\alpha}_{ip} \bar{\delta}_{ip}(x_i) sgn\bigg(\sum_{j\in N_i} k_{ij} \tilde{x}_{ij}\bigg)\,, 
\end{eqnarray}
\begin{eqnarray}
\label{eq:FTC:adaptivebounding_beforeIsolation}
\dot{\hat{\alpha}}_{ip} &=&  \Upsilon_{ip}\, \big|\sum_{j\in N_i} k_{ij} \tilde{x}_{ij}\big|\, \bar{\delta}_{ip}(x_i) \,, \\
\label{eq:FTC:adaptiveactuator_beforeIsolation}
\dot{\hat{\theta}}_{ip} &=& \cl{P}_{\bar{\theta}_{ip}} \bigg\{ \bar{\Gamma}_{ip} \sum_{j\in N_i} k_{ij} \tilde{x}_{ij} u_{ip} \bigg\}\,,
\end{eqnarray}
where 
$k_{ij}$ are positive constants, 
$\hat{\theta}_{ip}$ is an estimation of the actuator fault magnitude parameter $\theta_{ip}$ with the projection operator $\cl P$ restricting $\hat{\theta}_{ip}$ to the corresponding set $[\bar{\theta}_{ip},0]$ for $\bar{\theta}_{ip} \in (-1,0)$,
$\hat{\vartheta}_{ip}$ is an estimation of the neural network parameter vector $\vartheta_{ip}$, $\varphi_{ip} \udef col(\varphi_j : j=1,\cdots,\varrho)$ is the collective vector of fixed basis functions, $\hat{\alpha}_{ip}$ is an estimation of the unknown constants $\alpha_{ip}$, and $\Gamma_{ip}$ and $\Upsilon_{ip}$ are symmetric positive definite learning rate matrices.

Using some algebraic manipulations, we can rewrite (\ref{eq:FTC:controller_beforeIsolation}) as
$u_{ip} = \bar{u}_{ip}-\hat{\theta}_{ip} u_{ip}$. 
Therefore, using (\ref{eq:FTC:model_beforeIsolation}) and (\ref{eq:FTC:controller_beforeIsolation}), we can represent the collective closed-loop state dynamics as
\begin{equation}
\dot{x}^p = - \cl{L} x^p + \zeta^p - \bar{\zeta}^p + \tilde{f}^p + \delta^p -\psi^p+\varpi^p \,,
\label{eq:FTC:BeforeIsolation:dynamics2}
\end{equation}
where $x^p \in \Re^{M+1}$, $p=1,\cdots,n$, is comprised of the $p$th component of the $M$ agents and the leader as the $(M+1)$th agent, i.e., $x^p=\begin{bmatrix}
x_{1p}, & x_{2p}, & \cdots, & x_{Mp}, & x_p^r
\end{bmatrix}^T$, the terms $\zeta^p \in \Re^{M+1}$ and $\bar{\zeta}^p \in \Re^{M+1}$ are defined in (\ref{eq:eta_p}), and the terms $\tilde{f}^p \in \Re^{M+1}$, $\delta^p \in \Re^{M+1}$, $\psi^p \in \Re^{M+1}$ and $\varpi^p \in \Re^{M+1}$ are defined as
\begin{eqnarray}
\label{eq:f_tilde}
\tilde{f}^p &\udef& \begin{bmatrix} (\tilde{\vartheta}_{1p})^T \varphi_{1p} & \cdots & (\tilde{\vartheta}_{Mp})^T \varphi_{Mp} & 0
\end{bmatrix}^T \,,\\
\label{eq:delta}
\delta^p &\udef& \begin{bmatrix} \delta_{1p} &\cdots & \delta_{Mp} & 0  \end{bmatrix}^T \,,  \\
\label{eq:psi}
\psi^p &\udef& \begin{bmatrix} \psi_{1p} &\cdots & \psi_{Mp} & 0  \end{bmatrix}^T\,, \\
\label{eq:epsU}
\varpi^p &\udef& \begin{bmatrix} \tilde{\theta}_{1p} u_{1p} &\cdots & \tilde{\theta}_{Mp} u_{Mp} & 0  \end{bmatrix}^T
\,,
\end{eqnarray}
where $\tilde{\theta}_{ip}=\theta_{ip}-\hat{\theta}_{ip}$ is the actuator fault magnitude estimation error, and $\tilde{\vartheta}_{ip}=\vartheta_{ip}-\hat{\vartheta}_{ip}$ and $\varphi_{ip}$ are the parameter estimation errors and basis functions corresponding to the neural network model associated with the $p$th state component of the $i$th agent, respectively.
To derive  the adaptive algorithm and to investigate analytically the stability properties of the feedback system, we consider the following Lyapunov function candidate:
\begin{eqnarray}
\hspace{-5mm}V_p &=& {x^p}^T \Psi \, {x^p} + (\tilde{\vartheta}^{p})^T (\Gamma^p)^{-1}\tilde{\vartheta}^p + (\tilde{\alpha}^p)^T (\Upsilon^p)^{-1} \tilde{\alpha}^p \nonumber\\
&& + (\tilde{\theta}^{p})^T (\bar{\Gamma}^p)^{-1}\tilde{\theta}^p \,,
\label{eq:FTC:V_beforeIsolation}
\end{eqnarray}
where $\Psi$ is defined in Lemma~\ref{lem:laplaciansquared}, $\tilde{\vartheta}^p=\begin{bmatrix} \tilde{\vartheta}_{1p}^T, &\cdots, & \tilde{\vartheta}_{Mp}^T \end{bmatrix}^T$ is the collective parameter estimation errors, $\tilde{\alpha}^p=\begin{bmatrix} \tilde{\alpha}_{1p}, &\cdots, & \tilde{\alpha}_{Mp} \end{bmatrix}^T$ is the collective bounding parameter estimation errors defined as $\tilde{\alpha}_{ip}=\alpha_{ip}-\hat{\alpha}_{ip}$, $\tilde{\theta}^p=\begin{bmatrix} \tilde{\theta}_{1p}, &\cdots, & \tilde{\theta}_{Mp} \end{bmatrix}^T$ is the collective actuator fault magnitude parameter estimation errors, and $\Gamma^p={\rm diag} \{\Gamma_{1p},\cdots,\Gamma_{Mp}\}$, $\Upsilon^p={\rm diag} \{\Upsilon_{1p},\cdots,\Upsilon_{Mp}\}$ and $\bar{\Gamma}^p={\rm diag} \{\bar{\Gamma}_{1p},\cdots,\bar{\Gamma}_{Mp}\}$ are adaptive learning rate matrices.

Following the same procedure as given in the proof of Theorem~\ref{thm:nominal}, using (\ref{eq:f_tilde}), (\ref{eq:delta}), (\ref{eq:psi}) and (\ref{eq:epsU}), and selecting the adaptive algorithm for $\hat{\vartheta}_{ip}$ and $\hat \theta_{ip}$ as (\ref{eq:FTC:adaptive_beforeIsolation}) and (\ref{eq:FTC:adaptiveactuator_beforeIsolation}), respectively, it can be shown that the time derivative of the Lyapunov function (\ref{eq:FTC:V_beforeIsolation}) along the solution of (\ref{eq:FTC:BeforeIsolation:dynamics2}) satisfies
\begin{eqnarray*}
\dot{V}_p &\leq & - {x^p}^T \bar{\cl{L}} \, {x^p} \nonumber\\
&& +2 \sum_{i=1}^M \bigg( \sum_{j\in N_i} k_{ij} \tilde{x}_{ij} (\delta_{ip}-\psi_{ip}) - \tilde{\alpha}_{ip} (\Upsilon_{ip})^{-1} \dot{\hat{\alpha}}_{ip}\bigg) \,.
\label{eq:FTC:Vdot_beforeIsolation2}
\end{eqnarray*}
It is worth noting that since the projection modification can only make the Lyapunov function derivative more negative, the stability properties derived for the standard algorithm still hold (\cite{Polycarpou:book2006}).
By using (\ref{eq:FTC:adaptiveterm_beforeIsolation}) and based on Assumption~\ref{Assump:BoundingFunction}, we have
\begin{eqnarray} \label{eq:FTC1:bounding}
\lefteqn{\sum_{j \in N_i} k_{ij} \tilde{x}_{ij} (\delta_{ip}-\psi_{ip}) } \nonumber\\
&=& \sum_{j \in N_i} k_{ij} \tilde{x}_{ij} \big(\delta_{ip}-\hat{\alpha}_{ip} \bar{\delta}_{ip}  sgn(\sum_{j \in N_i} k_{ij} \tilde{x}_{ij}) \big) \nonumber\\
&\leq&  |\sum_{j \in N_i} k_{ij} \tilde{x}_{ij}| \tilde{\alpha}_{ip} \bar{\delta}_{ip} \,.
\end{eqnarray}
By Using (\ref{eq:FTC1:bounding}), we have
\begin{eqnarray*}
\dot{V}_p &\leq & - {x^p}^T \bar{\cl{L}} \, {x^p} +2 \sum_{i=1}^M \bigg( \big|\sum_{j\in N_i} k_{ij} \tilde{x}_{ij}\big| \tilde{\alpha}_{ip} \bar{\delta}_{ip} \nonumber\\
&& - \tilde{\alpha}_{ip} (\Upsilon_{ip})^{-1} \dot{\hat{\alpha}}_{ip}\bigg) \,.
\label{eq:FTC:Vdot_beforeIsolation2}
\end{eqnarray*}
Therefore, by using  (\ref{eq:FTC:adaptivebounding_beforeIsolation}) and after some algebraic manipulations, we have
\begin{eqnarray}
\hspace{-5mm}\dot{V}_p &\leq & - {x^p}^T \bar{\cl{L}} \, {x^p}=-2\sum_{i=1}^M \bigg(\sum_{j\in N_i} k_{ij} (x_{ip}-x_{jp})\bigg)^2\,.
\label{eq:FTC:Vdot_beforeIsolation3}
\end{eqnarray}
Based on the same reasoning logic as reported in the proof of Theorem~\ref{thm:nominal}, we conclude that $\dot{V}_p$ is negative semidefinite, and $x_{ip}-x_{jp}$, $\hat{\vartheta}_{ip}$ and $\hat{\alpha}_{ip}$ are uniformly bounded. Integrating both sides of (\ref{eq:FTC:Vdot_beforeIsolation3}), we know that $(x_{ip}-x_{jp}) \in L_2$. Since $(x_{ip}-x_{jp}) \in L_{\infty} \cap L_2$ and $\dot{x}_{ip}-\dot{x}_{jp} \in L_\infty$, based on Barbalat's Lemma, we can conclude that consensus is reached asymptotically, i.e., $x_{ip}-x_{jp}\rightarrow 0$ as $t\rightarrow \infty$. More specifically, $x_{ip}-x_p^r \rightarrow 0$ as $t\rightarrow \infty$ and therefore, the leader-follower consensus is reached asymptotically. 

The aforementioned design and analysis procedure is summarized as follows:
\begin{theorem} \label{thm:beforeIsolation}
Suppose that the bounding Assumption~\ref{Assump:BoundingFunction} holds. Then, if a fault is detected, the adaptive fault-tolerant law (\ref{eq:FTC:controller_beforeIsolation}), the weight parameter adaptive law (\ref{eq:FTC:adaptive_beforeIsolation}), the bounding parameter adaptive laws (\ref{eq:FTC:adaptiveterm_beforeIsolation}) and (\ref{eq:FTC:adaptivebounding_beforeIsolation}), and the actuator fault parameter adaptive law (\ref{eq:FTC:adaptiveactuator_beforeIsolation}) guarantee
\begin{enumerate}
\item all the signals and parameter estimates are uniformly bounded, i.e., $x_{ip}$, $\hat\vartheta_{ip}$ and $\hat\alpha_{ip}$ are bounded for all $t\in (T_i,T_{isol})$;
\item leader-follower consensus is achieved asymptotically with a time-varying reference state, i.e. $x_i-x^r\rightarrow 0 $ as $t \rightarrow \infty$.
\end{enumerate}
\end{theorem}


\subsection{Accommodation after Fault Isolation} \label{sec:FTC}
Let us now assume that the isolation procedure described in Section \ref{sec:Diagnosis} provides the information that fault $s$ has been isolated at time $T_{isol}$.
Based on the FTC architecture described by (\ref{eq:general:contr}), the controller is reconfigured again to further improve control performance based on the diagnostic information of isolated fault type. Below, we describe the cases of  process fault described by (\ref{eq:processfaultclass}) and  actuator fault given by (\ref{eq:actuatorfaultclass}), respectively. Without loss of generality, let the leader be the agent number
$M+1$ with a set of neighborhoods $N_{M+1}$.

%
{\em{Adaptive Fault-Tolerant Controller for Process Fault}}

After the isolation of the fault type $s$, i.e., $t\geq T_{isol}$, the dynamics of the system takes on the following form: for $p=1,\cdots,n$, 
\begin{equation} \label{eq:FTC:model2}
\begin{array}{ccl} 
\dot x_{ip} =\phi_{ip}(x_{ip}) + u_{ip} + \eta_{ip}(x_{i},t) + \theta_{ip}^s g_{ip}^s(x_i)\,. 
\end{array}
\end{equation}
The following adaptive fault-tolerant controller is adopted:
\begin{eqnarray}
\label{eq:FTC:controller}
u_{ip} &=& -\phi_{ip}(x_{ip}) - \sum_{j\in N_i} \big(k_{ij}\tilde{x}_{ij}\big)- \hat {\theta}_{ip}^T \,g_{ip}^s(x_{i})  \nonumber\\
&& - (\bar \eta_{ip} +\kappa_p) sgn\bigg(\sum_{j \in N_i} k_{ij} \tilde{x}_{ij}\bigg), \\  
\dot{\hat{\theta}}_{ip} &=& \, \Gamma_{ip} \sum_{j\in N_i} k_{ij} \tilde{x}_{ij} g_{ip}^s(x_{i})\,,
\label{eq:FTC:adaptive}
\end{eqnarray}
where
$\hat{\theta}_{ip}$ is an estimation of the unknown fault parameter vector, and $\Gamma_{ip}$ is a symmetric positive definite learning rate matrix. Then, we have the following:
\begin{theorem}
Assume that process fault $s$ occurs at time $T_i$ and that it is isolated at time $T_{isol}$. Then, the fault-tolerant controller (\ref{eq:FTC:controller}) and fault parameter adaptive law (\ref{eq:FTC:adaptive}) guarantee that
the leader-follower consensus is achieved asymptotically with a time-varying reference state, i.e., $x_i-x^r\rightarrow 0 $ as $t \rightarrow \infty$.
\end{theorem}
\begin{pf}
Based on (\ref{eq:FTC:model2}) and (\ref{eq:FTC:controller}), the closed-loop system dynamics are given by
\begin{eqnarray*}
\dot x_{ip} &=& -\sum_{j\in N_i} (k_{ij}\tilde{x}_{ij}) + \eta_{ip}(x_{i},t) + {\tilde{\theta}_{ip}}^T \,\, g_{ip}^s(x_{ip}) \nonumber\\
&& - (\bar \eta_{ip}+\kappa_p) sgn\bigg(\sum_{j \in N_i} k_{ij} \tilde{x}_{ij} \bigg)  \,. 
\label{eq:FTC:states}
\end{eqnarray*}
We can represent the collective output dynamics as
\begin{eqnarray}
\dot{x}^p = - \cl{L} x^p +\zeta^p - \bar{\zeta}^p + \tilde{f}^{sp}
\label{eq:FTC:dynamics2}
\end{eqnarray}
where $x^p \in \Re^{M+1}$, $p=1,\cdots,n$, is comprised of the $p$th component of the $M$ agents and the leader as the $(M+1)$th agent, i.e., $x^p=\begin{bmatrix}
x_{1p}, & x_{2p}, & \cdots, & x_{Mp}, & x_p^r
\end{bmatrix}^T$, the terms $\zeta^p \in \Re^{M+1}$ and $\bar{\zeta}^p \in \Re^{M+1}$ are defined in (\ref{eq:eta_p}), $\tilde{f}^{sp} \in \Re^{M+1}$ is defined as
\begin{eqnarray}
\label{eq:fs_tilde}
\tilde{f}^{sp} &\udef& \begin{bmatrix} \tilde{f}^s_{1p}, & \cdots, & \tilde{f}^s_{Mp}, & 0
\end{bmatrix}^T 
\end{eqnarray}
where $\tilde{f}^s_{ip} \udef (\tilde{\theta}_{ip})^T g_{ip}^s$, $i=1,\cdots,M$, and $\tilde{\theta}_{ip}=\theta_{ip}^*-\hat{\theta}_{ip}$ and $g_{ip}^s$ are the parameter estimation error and fault functions corresponding to the $p$th component of $M$ agents, respectively.
We consider the following Lyapunov function candidate:
\begin{eqnarray}
V_p &=& {x^p}^T \Psi x^p+ \tilde{\theta}^{pT} (\Gamma^p)^{-1} \tilde{\theta}^p \,,
\label{eq:FTC:V}
\end{eqnarray}
where $\Psi$ is defined in Lemma~\ref{lem:laplaciansquared}, $\tilde{\theta}^p=\begin{bmatrix} \tilde{\theta}_{1p}^T, &\cdots, & \tilde{\theta}_{Mp}^T \end{bmatrix}^T$ is the collective parameter estimation errors, and $\Gamma^p={\rm diag} \{\Gamma_{1p},\cdots,\Gamma_{Mp}\}$ is a positive definite adaptive learning rate matrix.
Then, using (\ref{eq:Nominal:Vdot2}), (\ref{eq:Nominal:Vdot3}), and the same reasoning logic for (\ref{eq:fs_tilde}), the time derivative of the Lyapunov function (\ref{eq:FTC:V}) along the solution of (\ref{eq:FTC:dynamics2}) is given by
\begin{eqnarray*}
\dot{V}_p &=& -{x^p}^T \bar{\cl{L}} \, {x^p}  + 2\sum_{i=1}^M \sum_{j\in N_i} k_{ij} \tilde{x}_{ij} (\eta_{ip}-\dot{x}_p^r) \nonumber \\
&& - 2\sum_{i=1}^M \sum_{j\in N_i} k_{ij} \tilde{x}_{ij} (\bar \eta_{ip}+\kappa_p) sgn\big(\sum_{j \in N_i} k_{ij} \tilde{x}_{ij}\big) \nonumber\\
&& + 2\sum_{i=1}^M \tilde{\theta}_{ip}^T \bigg(\sum_{j\in N_i} k_{ij} \tilde{x}_{ij} g_{ip}^s - (\Gamma_{ip})^{-1} \dot{\hat{\theta}}_{ip}\bigg)\,,
\label{eq:FTC:Vdot4}
\end{eqnarray*}
where $\bar{\cl{L}}$ is defined in (\ref{Lbar}). Therefore, 
choosing the adaptive law as (\ref{eq:FTC:adaptive}), we have
$\dot{V}_{p} \leq -{x^p}^T \bar{\cl{L}} \, {x^p}$.
Then, the proof can be concluded by using the same reasoning logic as reported in the analysis of Theorem~\ref{thm:beforeIsolation}.
\qed
\end{pf}
{\em{Adaptive Fault-Tolerant Controller for Actuator Fault}}

In the case of an actuator fault, i.e., $t\geq T_{isol}$, the dynamics of the system takes on the following form: for $p=1,\cdots,n$, 
\begin{equation} \label{eq:FTC:model2Actuator}
\begin{array}{ccl} 
\dot x_{ip} = \phi_{ip}(x_{ip}) + (1+\theta_{ip}) u_{ip} + \eta_{ip}(x_{i},t) \,.
\end{array}
\end{equation}
The following adaptive fault-tolerant controller is adopted:
\begin{eqnarray}
\label{eq:FTC:controllerActuatorAfter}
u_{ip} &=& \frac{1}{1+\hat{\theta}_{ip}} \bar{u}_{ip} \,, \\
\dot{\hat{\theta}}_{ip} &=& \cl{P}_{\bar{\theta}_{ip}} \bigg\{ \bar{\Gamma}_{ip} \sum_{j\in N_i} k_{ij} \tilde{x}_{ij} u_{ip}\bigg\} \,,
\label{eq:FTC:adaptiveActuatorAfter}
\end{eqnarray}
where $\bar{u}_{ip} \udef -\phi_{ip}(x_{ip}) - \displaystyle\sum_{j\in N_i} k_{ij}\tilde{x}_{ij} - \bar \kappa_{ip} sgn\big(\displaystyle \sum_{j \in N_i} k_{ij} \tilde{x}_{ij}\big)$,
$\hat{\theta}_{ip}$ is an estimation of the unknown actuator fault magnitude parameter $\theta_{ip}$ with the projection operator $\cl P$ restricting $\hat{\theta}_{ip}$ to the corresponding set $[\bar{\theta}_{ip},0]$ for $\bar{\theta}_{ip} \in (-1,0)$, and $\bar{\Gamma}_{ip}$ is a symmetric positive definite learning rate matrix. Then, we have the following:
\begin{theorem}
Assume that an actuator fault occurs at time $T_i$ and that it is isolated at time $T_{isol}$. Then, the fault-tolerant controller (\ref{eq:FTC:controllerActuatorAfter}) and fault parameter adaptive law (\ref{eq:FTC:adaptiveActuatorAfter}) guarantee that
the leader-follower consensus is achieved asymptotically with a time-varying reference state, i.e., $x_i-x^r\rightarrow 0 $ as $t \rightarrow \infty$;
\end{theorem}
\begin{pf}
Using some algebraic manipulations, we can rewrite (\ref{eq:FTC:controllerActuatorAfter}) as
$u_{ip} = \bar{u}_{ip}-\hat{\theta}_{ip} u_{ip}$. Therefore, substituting $u_{ip}$ in (\ref{eq:FTC:model2Actuator}), the closed-loop system dynamics are given by
\begin{eqnarray*}
\dot x_{ip} &=& -\sum_{j\in N_i} (k_{ij}\tilde{x}_{ij}) + \eta_{ip}(x_{i},t) +\tilde{\theta}_{ip} u_{ip}  \nonumber\\
&& - (\bar \eta_{ip}+\kappa_p) sgn\bigg(\sum_{j \in N_i} k_{ij} \tilde{x}_{ij} \bigg) \,. 
\label{eq:FTC:statesActuator}
\end{eqnarray*}
We can represent the collective output dynamics as
\begin{eqnarray}
\dot{x}^p = - \cl{L} x^p + \zeta^p - \bar{\zeta}^p + \varpi^p
\label{eq:FTC:dynamics2Actuator}
\end{eqnarray}
where $x^p \in \Re^{M+1}$, $p=1,\cdots,n$, is comprised of the $p$th component of the $M$ agents and the leader as the $(M+1)$th agent, i.e., $x^p=\begin{bmatrix}
x_{1p}, & x_{2p}, & \cdots, & x_{Mp}, & x_p^r
\end{bmatrix}^T$, and the terms $\zeta^p \in \Re^{M+1}$, $\bar{\zeta}^p \in \Re^{M+1}$ and $\varpi^p \in \Re^{M+1}$ are defined in (\ref{eq:eta_p}) and (\ref{eq:epsU}).

We consider the following Lyapunov function candidate:
\begin{eqnarray}
V_p &=& {x^p}^T \Psi x^p+ \tilde{\theta}^{pT} (\bar{\Gamma}^p)^{-1} \tilde{\theta}^p \,,
\label{eq:FTC:VActuator}
\end{eqnarray}
where $\Psi$ is defined in Lemma~\ref{lem:laplaciansquared}, $\tilde{\theta}^p=\begin{bmatrix} \tilde{\theta}_{1p}, &\cdots, & \tilde{\theta}_{Mp} \end{bmatrix}^T$ is the collective actuator fault magnitude parameter estimation errors, and $\bar{\Gamma}^p={\rm diag} \{\bar{\Gamma}_{1p},\cdots,\bar{\Gamma}_{Mp}\}$ is a positive definite adaptive learning rate matrix.
Then, using (\ref{eq:Nominal:Vdot2}) and (\ref{eq:Nominal:Vdot3}), and the same reasoning logic for (\ref{eq:epsU}), the time derivative of the Lyapunov function (\ref{eq:FTC:VActuator}) along the solution of (\ref{eq:FTC:dynamics2Actuator}) is given by
\begin{eqnarray*}
\dot{V}_p &=& -{x^p}^T \bar{\cl{L}} \, {x^p}  + 2\sum_{i=1}^M \sum_{j\in N_i} k_{ij} \tilde{x}_{ij} (\eta_{ip}-\dot{y}_p^r) \nonumber\\
&&+ 2\sum_{i=1}^M \tilde{\theta}_{ip} \bigg(\sum_{j\in N_i} k_{ij} \tilde{x}_{ij} u_{ip} - (\bar{\Gamma}_{ip})^{-1} \dot{\hat{\theta}}_{ip}\bigg) \nonumber \\
&& - 2\sum_{i=1}^M \sum_{j\in N_i} k_{ij} \tilde{x}_{ij} (\bar \eta_{ip}+\kappa_p) sgn\big(\sum_{j \in N_i} k_{ij} \tilde{x}_{ij}\big)\,,
\label{eq:FTC:Vdot4Actuator}
\end{eqnarray*}
where $\bar{\cl{L}}$ is defined in (\ref{Lbar}). Therefore, 
choosing the adaptive law as (\ref{eq:FTC:adaptiveActuatorAfter}), we have
$\dot{V}_{p} \leq -{x^p}^T \bar{\cl{L}} \, {x^p}$.
Then, the proof can be concluded by using the same reasoning logic as reported in the analysis of Theorem~\ref{thm:beforeIsolation}.
\qed
\end{pf}

\section{Simulation Results} \label{sec:simulation}
In this section, a simulation example of a networked multi-agent system consisting of 5 agents is considered to illustrate the effectiveness of the distributed fault-tolerant control method.
The dynamics of each agent is given by
\begin{equation}
\begin{split}
\dot{x}_{i} = u_{i} + \eta_i + \beta_i(t-T_i) f_i(x_i,u_i) \\
\end{split} \,,
\label{eq:simulation-model}
\end{equation}
where, for $i=1,\cdots,5$, $x_i=[x_{i1}, x_{i2}]^T$ and $u_i=[\bar \nu_i cos(\bar{\psi}_i), \bar \nu_i sin(\bar{\psi}_i)]^T$ are the state and input vector of $i$th agent, respectively, $\bar{\psi}_i$ and $\bar \nu_i$ in the input vector $u_i$ are the orientation and the linear velocity of each agent representing a ground vehicle.

The ground vehicle given model in (\ref{eq:simulation-model}) is a standard unicycle-like model that can be controlled with the orientation $\bar \psi_i$ and vehicle linear velocity $\bar \nu_i$. Using the developed algorithms, the desired orientation and linear velocity of the ground vehicle robot can be obtained uniquely. Then, a low level controller can be designed to track the desired orientation and linear velocity for driving the ground vehicles to desired positions.

The unknown modeling uncertainty in the local dynamics of the agents are assumed to be sinusoidal signals $\eta_i = [0.5 sin(t), 0.5 sin(t)]^T$ bounded by $\bar{\eta}_{i}=[0.6, 0.6]^T$. 
The objective is for each agent to follow the leader's position described by $x^r=[x_1^r,  x_2^r]^T=[5+sin(t), 5+cos(t)]^T$.   

The Laplacian matrix of the intercommunication graph of agents plus leader is given as
\begin{equation*}
\cl{L}  = \begin{bmatrix}
  \hspace{2mm}2\, & -1\, & \hspace{2mm}0\, & \hspace{2mm}0\, & -1\, & \hspace{2mm}0  \\
  \\
  -1\, & \hspace{2mm}3\, & \hspace{2mm}0\, & \hspace{2mm}0\, & -1\, & -1  \\
  \\
  \hspace{2mm}0\, & \hspace{2mm}0\, & \hspace{2mm}2\, & -1\, & -1\, & \hspace{2mm}0  \\
  \\
  \hspace{2mm}0\, & \hspace{2mm}0\, & -1\, & \hspace{2mm}2\, & -1\, & \hspace{2mm}0  \\
  \\
  -1\, & -1\, & -1\, & -1\, & \hspace{2mm}4\, & \hspace{2mm}0  \\
  \\
  \hspace{2mm}0\, & \hspace{2mm}0\, & \hspace{2mm}0\, & \hspace{2mm}0\, & \hspace{2mm}0\, & \hspace{2mm}0  \\
 \end{bmatrix}\,.
\end{equation*}
The fault class under consideration is defined as follows
\begin{enumerate}
\item A process fault described by $f_i^1=\theta_i^1 g_i^1$, where $g_i^1=x_i^2 cos(x_i)$ and the fault magnitude $\theta_i^1 \in [0, 1]$. 
\item An actuator fault described by $f_i^2=\theta_i^2 g_i^2$, where $g_i^2=u_i$ and the fault magnitude $\theta_i^2 \in [-0.8,0]$. 
\end{enumerate}

The observer gain for fault detection estimator is chosen as $h_i=2$. For fault isolation estimator $\lambda_i=10$ has been chosen. Based on the magnitude of the fault types, we choose the center and radius of the parameter projection sphere as $O_i^1=0.5$ and $R_i^1=0.5$ for the process fault type and $O_i^2=-0.4$ and $R_i^1=0.4$ for the actuator fault type, respectively.

A radial basis function (RBF) neural network is used for approximation of the process fault function after its detection and before its isolation. The RBF network considered in this paper consists of 21 neurons with 21 adjustable gain parameters. The center of radial basis functions are equally distributed on interval $[-10,10]$ with a variance of 0.5. The initial parameter vector of the neural network is set to zero. We set the learning rate as $\Gamma_i=5$ and consider a constant bound on the network approximation error, i.e., $\bar{\delta}_i=1$. The adaptive gains in (\ref{eq:FTC:adaptivebounding_beforeIsolation}) and (\ref{eq:FTC:adaptiveactuator_beforeIsolation}) are chosen as $\Upsilon_i=2$ and $\bar{\Gamma}_i=1$, respectively.

After fault isolation, the controller is reconfigured to accommodate the specific fault that has been isolated. We set the adaptive gain $\Gamma_i=0.2$ with a zero initial condition (see (\ref{eq:FTC:adaptive})).

\begin{figure}[thpb]
	  \centering
      \includegraphics[width= 7.5 true cm] {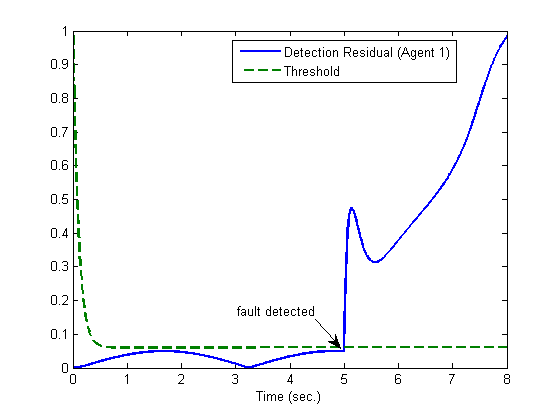}
      \caption{The case of a process fault in agent 1: fault detection residuals (solid and blue line) and the corresponding threshold (dashed and green line) generated by the FDE of agent 1
      \label{fig:pf_FDE1}}
\end{figure}

\begin{figure}
      \centering
      \includegraphics[width= 7.5 true cm] {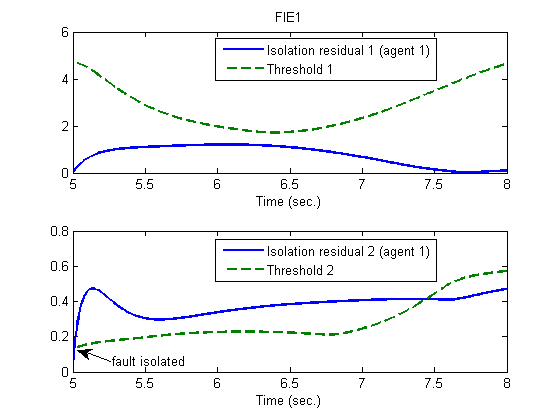}
      \caption{The case of a process fault in agent 1: the fault isolation residuals (solid and blue line) associated with two types of faults and the corresponding threshold (dashed and green line) generated by the FIEs of agent 1
      \label{fig:pf_FDE2}}
\end{figure}

Figure~\ref{fig:pf_FDE1} and  Figure~\ref{fig:pf_FDE2}
show the fault detection and isolation results when the first process fault class (i.e., $f_1^1=\theta_1^1 g_1^1$) with a magnitude of 0.8 occurs to agent 1 at $T_i = 5$ second. As can be seen from Figure~\ref{fig:pf_FDE1}, the residual corresponding to the output generated by the local FDE designed for agent 1 exceeds its threshold immediately after fault occurrence. Therefore, the process fault in agent 1 is timely detected. It can be seen in Figure~\ref{fig:pf_FDE2} that the residual corresponding to the FIE associated with the first fault type always remains below the threshold, while the residual corresponding with the FIE associated with the second fault type exceeds the threshold immediately after fault occurrence. Thus, based on the fault isolation decision scheme described in section \ref{subsec:Isolation}, the occurrence of fault type 1 can be concluded. The fault diagnosis results for the second states have the same behavior and therefore are omitted.

\begin{figure}[thpb]
	  \centering
      \includegraphics[width= 7 true cm] {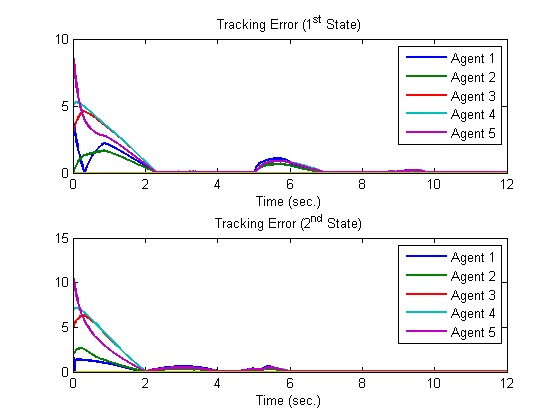}
      \caption{The tracking errors in the case of a process fault in agent 1: with adaptive fault-tolerant controllers
      \label{fig:pf_control}}
      
\end{figure}

\begin{figure}[thpb]
      \centering
      \includegraphics[width= 7 true cm] {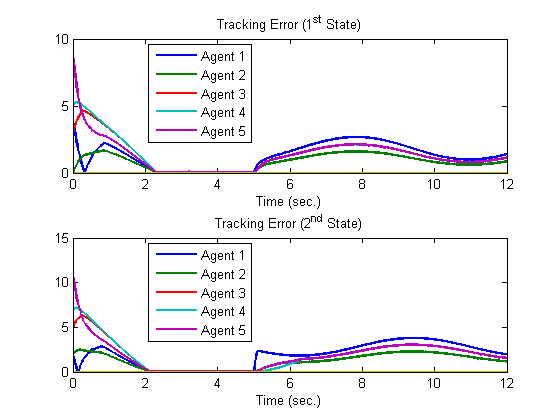}
      \caption{The tracking errors in the case of a process fault in agent 1: without adaptive fault-tolerant controllers
      \label{fig:pf_nocontrol}}
      
\end{figure}

In Figure~\ref{fig:pf_control} and Figure~\ref{fig:pf_nocontrol} we compare the leader-following performance of the agents under the action of the proposed adaptive FTCs. 
Regarding the performance of the adaptive fault-tolerant controllers, as can be seen from Figure~\ref{fig:pf_control},
the leader-following consensus is achieved using the proposed adaptive FTCs, while the agents cannot follow the leader
 without the FTC controllers after fault occurrence (see Figure~\ref{fig:pf_nocontrol}).
Thus the benefits of the FTC method can be clearly seen. 

\section{Conclusion}

In this paper, we investigate the problem of a distributed FTC for a class of multi-agent uncertain systems. Under certain assumptions, adaptive FTC controllers are developed to achieve the leader-following consensus in the presence of faults. The extensions to systems with more general structure is an interesting topic for future researches.


\bibliography{MyReferences}             
                                                   
%

\end{document}